\documentclass[reprint,
superscriptaddress,
amsmath,amssymb,
aps,
pra,
]{revtex4-2}

\usepackage[T1]{fontenc}
\usepackage[utf8]{inputenc}

\usepackage{graphicx}
\usepackage{dcolumn}
\usepackage{bm}
\usepackage{hyperref}

\usepackage{inputenc}
\usepackage{calc, graphicx, amsthm, geometry}
\usepackage{xcolor,multirow, tikz}
\usepackage{hyperref, cleveref, url}
\usepackage{amsfonts}
\usepackage{bbold}
\usepackage{blkarray}
\usepackage{latexsym}
\usepackage{mathrsfs}
\usepackage{thmtools}
\usepackage{mathtools}
\usepackage{pgfplots}
\usepackage{physics}
\usepackage{cancel}
\usepackage{tcolorbox} 

\usepackage{caption}
\usepackage{subcaption}

\begin{document}

\preprint{APS/123-QED}

\title{Casimir repulsion turned into attraction by the nonlocal response of salted water}

\author{Larissa Inácio}
 \email{larissa.inacio@live.com}
 \affiliation{Instituto de F\'{\i}sica, Universidade Federal do Rio de Janeiro \\ Caixa Postal 68528,   Rio de Janeiro,  Rio de Janeiro, 21941-972, Brazil}
 \author{Felipe S. S. Rosa}
 \email{frosa@if.ufrj.br}
\affiliation{Instituto de F\'{\i}sica, Universidade Federal do Rio de Janeiro \\ Caixa Postal 68528,   Rio de Janeiro,  Rio de Janeiro, 21941-972, Brazil}
\author{Serge Reynaud}
\email{serge.reynaud@lkb.upmc.fr}
\affiliation{Laboratoire Kastler Brossel, Sorbonne Université, CNRS, ENS-PSL, Collège de France, Campus Jussieu, F-75252 Paris, France}
\author{Paulo A. Maia Neto}
 \email{pamn@if.ufrj.br}
\affiliation{Instituto de F\'{\i}sica, Universidade Federal do Rio de Janeiro \\ Caixa Postal 68528,   Rio de Janeiro,  Rio de Janeiro, 21941-972, Brazil}%

\date{\today}%

\begin{abstract}
The Dzyaloshinskii-Lifshitz-Pitaevskii (DLP) theory of Casimir  forces predicts a repulsion between two material surfaces separated by a third medium with an intermediate dielectric function. This DLP repulsion paradigm constitutes an important example with many applications. We show here that it is broken when the surfaces interact across salted water due to the nonlocal response of the ions in solution. 
We consider the interaction between silica and gold and show that the zero-frequency contribution, which is the only one affected by the nonlocal response, is dominant for distances in the range near $100\,{\rm nm}$ and beyond. As a result, the total Casimir force between gold and silica surfaces in salted water is always attractive in spite of the DLP repulsion paradigm.
\end{abstract}

\maketitle

\section{Introduction}

The electromagnetic Casimir force \cite{Casimir1948,Decca2011,Woods2016,Gong2021} is usually 
 attractive. In a landmark work, however, it was shown by Dzyaloshinskii, Lifshitz and Pitaevskii (DLP) \cite{Dzyaloshinskii1961} that it can become  repulsive  when 
 the interacting surfaces with dielectric functions
$\epsilon_1$ and $\epsilon_2$ are separated by a liquid 
medium with dielectric function $\epsilon_3$ 
satisfying the so-called staircase condition $\epsilon_1< \epsilon_3<\epsilon_2$ 
over the relevant spectral range. 
Several materials satisfying such a condition were analyzed~\cite{VanZwol2010} and experimental observations of the repulsion  were reported in the  non-retarded short-distance van der Waals regime \cite{Milling1996,Lee2002} as well as in the retarded regime~\cite{Munday2009}.
Important applications have been implemented and proposed~\cite{Esteso2024}:
the thickening of a superfluid helium film on the surface of a container~\cite{Sabisky1973,Parsegian2005,Israelachvili2011}, the reduction of friction in nanoscale devices~\cite{Feiler2008}, and quantum levitation~\cite{Esteso2015}, for instance 
with a magnetodielectric plate~\cite{Inui2012, Shelden2023}.
The repulsive component of the Casimir force would provide a fine tuning of the distance in Fabry-P\'erot nanocavities~\cite{Ge2024}. 

In this paper, we show that the DLP repulsion paradigm for materials  separated by a medium satisfying the DLP staircase configuration can be broken, with the Casimir force turning back to attractive due to the nonlocal response of the immersion medium. 

Recent experiments have probed the Casimir force~\cite{munkhbat2021tunable} and torque~\cite{Kucucoz2024} between two gold nanoflakes in aqueous solution. The attractive electromagnetic Casimir force between a gold nanoflake and a metallic surface was shown to counteract a repulsive critical Casimir interaction in a binary liquid mixture~\cite{schmidt2023tunable}. 
Here we discuss the case of a gold nanoflake in salted water above a silica planar bulk surface, so that the DLP staircase condition is satisfied at the limit of zero frequency. In this case, 
the thermal contribution is dominant at distances larger than $\sim 100\,{\rm nm}.$ We show that the zero-frequency contribution as well as the total force are attractive at all length scales, due to the change of sign of the
reflection amplitude at the water-metal interface
when the nonlocal dielectric response in salted water is taken into account. 

The nonlocal response in metals leads to a noticeable modification of the Casimir-Polder interaction in the case of sharp edges~\cite{Kristensen2023}, but the corrections to the  force are typically small for graphene \cite{Rodriguez-Lopez2024} as well as for metallic surfaces~\cite{Esquivel-Sirvent2006,Hannemann2021}. In contrast, the nonlocal response in electrolytes leads to a change in the nature of the interaction, from repulsion to attraction, in the configuration analyzed in the present paper. 

{The change from repulsion into attraction by adding salt was  found in binary mixtures as a result of the interplay between the double-layer electrostatic and critical Casimir forces~\cite{Bier2011}. In contrast, we assume that the electromagnetic Casimir interaction is independent of the charge densities responsible for the electrostatic repulsion. }
We generalize the scattering formalism for electrolyte solutions developed in Ref.~\cite{maia2019scattering} in order to consider the case where one of the dielectric surfaces is replaced by a metallic one. {Our approach is based on the hydrodynamical model for the ions in solution considered within the bulk approximation~\cite{davies1972van}.
  It} allows us to 
  isolate the longitudinal and transverse zero-frequency contributions. While the former is repulsive and screened over the Debye length $\lambda_{\rm D}$, the latter provides an unscreened, attractive and larger contribution.
  Thus, the total force is attractive for all distances and values of $\lambda_{\rm D}.$
  {Note that corrections to the bulk approximation would be expected at very short distances. }
    
The paper is organized as follows. In Sec.~II, we discuss the parameters of the configuration which is analyzed within the scattering formalism in Sec.~III.  The  reflection coefficient for the gold-salted water interface is derived in Sec.~IV. Results for the Casimir interaction and concluding remarks  are presented in Secs.~V and VI, respectively.

\section{Gold nanoflake above a  silica surface}

\begin{figure}[h]
    \centering
    \includegraphics[width=0.7\linewidth]{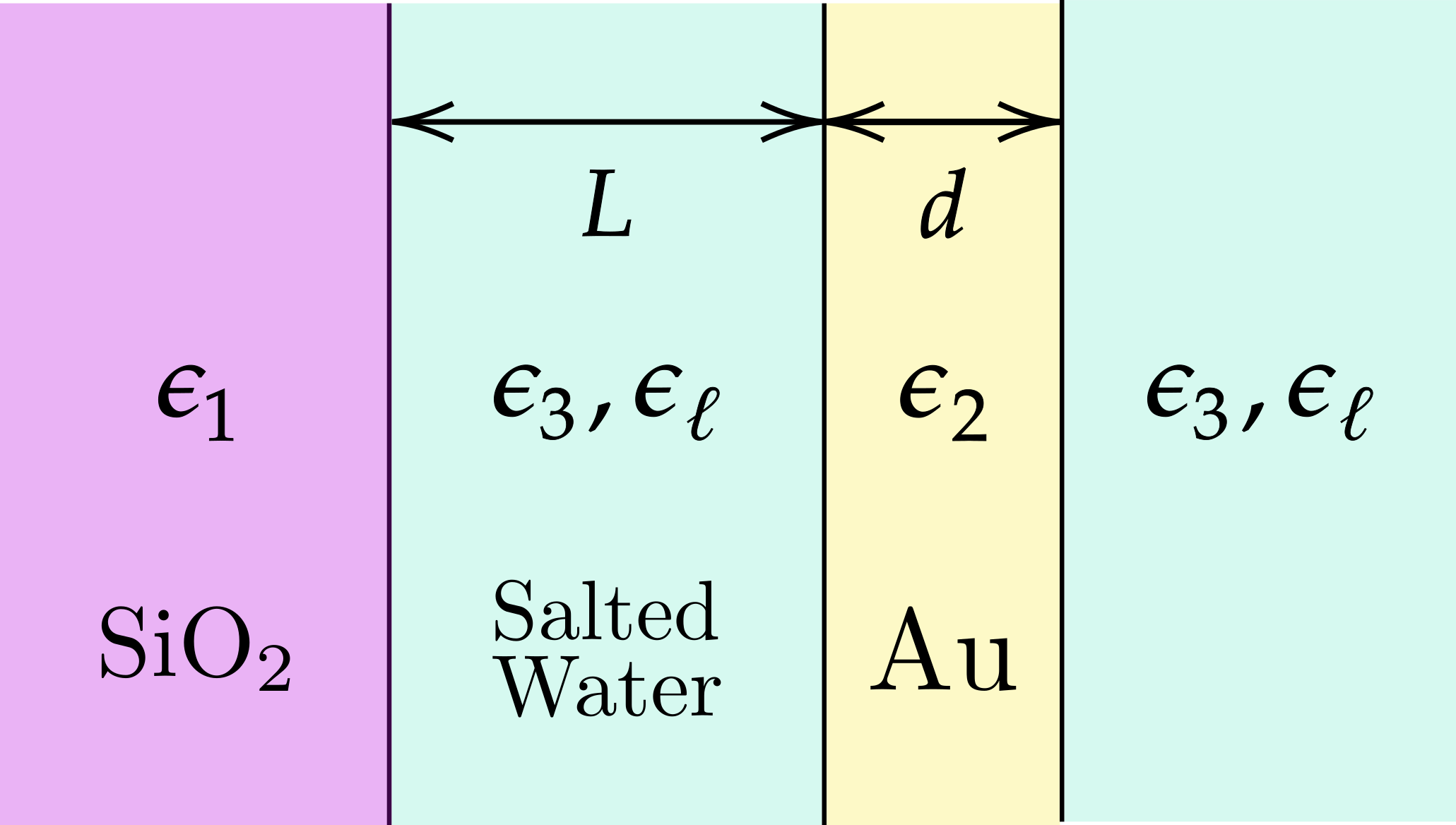}
\caption{Configuration with four regions and their corresponding dielectric functions: a half space of SiO$_2$ ($\epsilon_1$), a layer of salted water ($\epsilon_3,\epsilon_\ell$)  and a slab of gold ($\epsilon_2$) of thicknesses $L$ and $d,$ respectively.}
    \label{fig:configuration}
\end{figure}

In this work we investigate the setup depicted in Fig.~\ref{fig:configuration}. The half-space on the left is made of SiO$_2$, characterized by a dielectric function $\epsilon_1(i\xi)$ of imaginary frequencies $i\xi,$ which is described by a Drude-Lorentz model with the parameters
of Ref.~\cite{VanZwol2010} (more specifically set 1 in that reference). The SiO$_2$ surface interacts with the gold nanoflake modelled as a slab 
 of thickness $d$
 {across a distance $L$ in a region filled with salted water.
 The dielectric  function of gold}
$\epsilon_2(i\xi)$ is obtained from optical data~\cite{Palik1998} at real frequencies by using a dispersion relation~\cite{lambrecht2000casimir}.
 We complete the data at low frequencies by taking the Drude model
 \begin{equation}
    \epsilon_{\rm Dr}(i\xi) = 1 + \dfrac{\omega_{p2}^2}{\xi(\xi + \gamma_2)}
\end{equation}
 with parameters $\omega_{p2}=9\,{\rm eV}$ and $\gamma_2=35\,{\rm meV}.$ 
 
Remarkably, the main conclusion of the paper does not depend on the model employed for describing the metallic response. Repulsion would also be turned into attraction if gold was described by the dissipationless plasma model.
{In addition, the change from repulsion to attraction does not depend on the thickness $d.$
Although the original DLP theoretical framework~\cite{Dzyaloshinskii1961} was developed for two semi-infinite half-spaces, the staircase condition might also lead to repulsion in the case of finite-thickness slabs~\cite{Zhao2011}. Indeed, the modification of the Casimir interaction with respect to a half-space is  small~\cite{Esquivel-Sirvent2008,Pirozhenko2008} for the thicknesses ($d\sim 50\,{\rm nm}$) and distances ($L\sim 100\,{\rm nm}$)   of experimental relevance~\cite{munkhbat2021tunable}. We take nonetheless the finite thickness into account in the calculations presented below.}

We take salted water as the immersion medium. The presence of ions in solution gives rise to a nonlocal tensorial electromagnetic response.
{Within the bulk approximation, the interacting surfaces are not taken into account at the level of the constitutive equations.  
The resulting dielectric tensor in reciprocal space ~\cite{davies1972van} %
\begin{align}
\boldsymbol{\epsilon}_{\rm SW}(i\xi,K)= \begin{pmatrix}
\epsilon_{3}(i\xi)& 0 & 0 \\
0 & \epsilon_{3}(i\xi)  & 0  \\
0 & 0  & \epsilon_{\ell}(i\xi,K) 
\end{pmatrix}
\end{align}
therefore describes the response of an homogeneous electrolyte solution. 
In the transverse subspace we have}
\begin{align}
  \epsilon_{3}(i\xi) &=  \epsilon_b(i\xi) + \frac{\omega_{p3}^2}{\xi(\xi + \gamma_3)} \label{epsilon1}
  \end{align}
  while the longitudinal component reads
  \begin{align}
   \epsilon_{\ell}(i\xi,K) &=  \epsilon_b(i\xi) +\bigg( \frac{\xi(\xi +\gamma_3)}{\omega_{p3}^2}+\frac{\lambda_{\rm D}^2}{\epsilon_{b0}}K^2\bigg)^{-1}\label{epsilonlongi}.
\end{align}
Here, $\epsilon_b(i\xi)$ is the dielectric function of pure water, described by a Lorentz model with parameters given by Ref.~\cite{VanZwol2010}. 

In addition to $\epsilon_b(i\xi),$ the transverse dielectric function  $\epsilon_{3}(i\xi)$ contains a Drude term accounting for the ionic current, which is
given in terms of a plasma frequency
$\omega_{p3}$ and a relaxation parameter $\gamma_3.$
The plasma frequency $\omega_{p3}$ is several orders of magnitude smaller than its gold counterpart $\omega_{p2}$ and the Drude term in 
(\ref{epsilon1}) is negligible at all nonzero Matsubara frequencies.
The transverse dielectric functions
of all three materials shown in Fig.~\ref{fig:configuration} are plotted in Fig.~\ref{fig:permitivitties}.

The nonlocal response of the ionic current is captured by the $K$ dependence of the longitudinal 
dielectric component given by Eq.~(\ref{epsilonlongi}). The characteristic nonlocal 
length scale is set by $\lambda_{\rm D}/\sqrt{\epsilon_{b0}}$
 where
 the Debye screening length $\lambda_{\rm D}$ 
 for monovalent ions at temperature $T$
 is given 
 in terms of the ionic concentration $N$ by
\begin{align}
\lambda_{\rm D}=\sqrt{\frac{\epsilon_{b0}k_{\rm B}T}{N e^2}}
\end{align}
where  $\epsilon_{b0}= \epsilon_b(0),$ $k_{\rm B}$ is the Boltzmann constant, and $e$ is the elementary charge.  

The spatially-dispersive $\epsilon_{\ell}(i\xi,K)$ allows for longitudinal modes, which contribute to the Casimir interaction as discussed in the next section.


\section{Scattering formalism for the Casimir interaction} \label{scattering-section}

Within the scattering formalism \cite{lambrecht2006casimir,rahi2009scattering}, we  write the Casimir interaction free energy in terms of the roundtrip operator $\mathcal{M}(\abs{\xi_n})$ as follows:
\begin{eqnarray}
&&\mathcal{F}_{\rm cas}=\sum_{n=-\infty}^\infty \mathcal{F}_n    \\
&&\mathcal{F}_n=\frac{k_{\rm B}T}{2}\int \frac{d^2k}{(2\pi)^2} \log \det \big(1-\mathcal{M}(\abs{\xi_n})\big) \label{logdet} 
\\
&&\mathcal{M}(\abs{\xi_n})=\mathcal{R}_1\mathcal{T}_{12}\mathcal{R}_2\mathcal{T}_{21} \label{opM}  
\end{eqnarray}
where $\xi_n = 2 \pi n k_{\rm B}T/\hbar$ are the Matsubara frequencies. The reflection  $\mathcal{R}_i$ and translation $\mathcal{T}_{ij}$ operators are represented as matrices accounting for the two transverse polarizations and the longitudinal one in the plane wave basis~\cite{maia2019scattering}. The two interacting surfaces 
are labeled $i=1,2$
for the dielectric-water and the water-nanoflake interfaces, respectively (see 
Fig.~\ref{fig:configuration}). 
The translation matrices are diagonal and read 
\begin{align}    
\mathcal{T}_{12}= \mathcal{T}_{21}= \begin{pmatrix}
e^{-\kappa_3 L} & 0 & 0 \\
0 & e^{-\kappa_3 L}  & 0 \\
0 & 0 & e^{-\kappa_\ell L} \\
\end{pmatrix}  
\end{align}
where 
\begin{eqnarray}
&&\kappa_3 = \sqrt{k^2 + \epsilon_3(i\xi) \, \frac{\xi^2}{c^2}} \\
&&\kappa_{\ell} = \sqrt{k^2 + \frac{1}{v_{\rm th}^2}\left[\xi(\xi+\gamma_3) + \frac{\omega_{p3}^2}{\epsilon_b(i\xi)} \right]}\label{kappaell}
\end{eqnarray}
with $v_{\rm th} = \sqrt{k_{\rm B} T/m}$ representing the {characteristic scale of the thermal ion velocity ($m$ denotes the mass of the ions)~\cite{maia2019scattering}. }

The reflection matrices are block-diagonal:
\begin{align}
\mathcal{R}_{i}= \begin{pmatrix}
r_{ss}^{(i)}& 0 & 0 \\
0 & r_{pp}^{(i)}  & r_{p \ell}^{(i)}  \\
0 & r_{\ell p}^{(i)}  & r_{\ell \ell}^{(i)} \\
\end{pmatrix}
\end{align}
where $s,p,\ell$ denote, respectively, the transverse electric (TE), transverse magnetic (TM) and longitudinal modes. 
The reflection amplitudes $r^{(i)}_{\alpha \beta}$ are determined  from the continuity of tangential components of the electric and magnetic fields, together with the condition that the normal component of the ionic current vanishes at the interface.  Information about whether the surface corresponds to a silica half-space ($i=1$)  or a gold slab ($i=2$) is encoded in $r_{\alpha\beta}^{(i)}$.
 
The behavior of such amplitudes are considerably different for zero ($n=0$) and finite ($n \neq 0$) Matsubara frequencies, so it is necessary to treat these instances separately.
For finite frequencies, 
the contribution from the longitudinal channel is exponentially suppressed over an extremely small length scale as 
$\kappa_{\ell}\ge 2\pi\sqrt{mk_{\rm B}T}/\hbar$ scales as the inverse thermal de Broglie wavelength of the ions.
In addition, the Drude term in (\ref{epsilon1}) is negligible at nonzero Matsubara frequencies.
 As a result, the contribution from finite frequencies is exactly as if there was no electrolyte and reads
\begin{widetext}
\begin{equation}
   \mathcal{F}_{n} = \frac{k_{\rm B}T}{2}
 \sum_{\alpha =s,p} \int \frac{d^2k}{(2\pi)^2} \log \big[1- r_{\alpha\alpha}^{(1)}(i|\xi_n|) r_{\alpha\alpha}^{(2)}(i|\xi_n|) e^{-2\kappa_3 L} \big]. \quad\quad (n\neq 0) 
   \label{finite} 
\end{equation}
\end{widetext}
For finite frequencies, $r_{\alpha\alpha}^{(1)}(i|\xi_n|)$ and 
$r_{\alpha\alpha}^{(2)}(i|\xi_n|)$
are the usual Fresnel coefficients for a \ half-space and a slab~\cite{Esquivel-Sirvent2008,Pirozhenko2008}, respectively, both considered within a local theory. For convenience, the latter is discussed in Appendix~\ref{sec-coefficients}. 

On the other hand, the zero-frequency contribution is strongly modified by the ions in solution. 
In the limit $\xi \rightarrow 0$,
the diagonal TE reflection amplitude vanishes as well as 
the contribution from nondiagonal elements of $\cal R$  since
\begin{align}
    \lim_{\xi \rightarrow 0} r_{p \ell}^{(i)}(i\xi) \,r_{\ell p}^{(j)}(i\xi) = 0. \hspace{15pt} (i,j=1,2;\,i\neq j)
\end{align}
Thus,  we obtain separate contributions associated with TM and longitudinal polarizations:
\begin{align}\label{F0def}
    \mathcal{F}_0=&\mathcal{F}_0^\text{\scriptsize TM}+\mathcal{F}_0^\text{\scriptsize long}
\end{align}
\begin{align} \label{F0TM}
\mathcal{F}_0^\text{\scriptsize TM} =&\frac{k_{\rm B}T}{2}  \int \frac{d^2k}{(2\pi)^2} \log \big[1- r_{pp}^{(1)}(0)\, r_{pp}^{(2)}(0)\, e^{-2 k L} \big] 
    \end{align}
\begin{align} \label{F0long}
    \mathcal{F}_0^\text{\scriptsize long} =& \frac{k_{\rm B}T}{2}  \int \frac{d^2k}{(2\pi)^2} \times \\\nonumber 
   &\log \big[1- r_{\ell\ell}^{(1)}(0)\, r_{\ell\ell}^{(2)}(0) \, e^{-2 \sqrt{k^2 + 1/\lambda_{\rm D}^2} L} \big] 
\end{align}
The reflection coefficients
for the SiO$_2$-water interface are~\cite{maia2019scattering} 
 $r_{pp}^{(1)}(0)=-1$ and
\begin{equation}\label{rellell}
    r_{\ell\ell}^{(1)}(0)=\frac{1-\frac{k}{\kappa_\ell}\frac{\epsilon_{1}(0)}{\epsilon_{b0}}}{1+\frac{k}{\kappa_\ell}\frac{\epsilon_{1}(0)}{\epsilon_{b0}}}.
\end{equation}
From (\ref{kappaell}) we have $\kappa_{\ell}>k$, and
since $\epsilon_{b0}>\epsilon_{1}(0),$
we have 
$r_{\ell\ell}^{(1)}(0)>0.$

The reflection coefficients $r_{pp}^{(2)}(0)$ and $r_{\ell\ell}^{(2)}(0)$ for the water-nanoflake interface are discussed in Appendix~\ref{sec-coefficients}.
It turns out that they are independent of the nanoflake thickness and, therefore, equal to the expressions for a semi-infinite gold half-space in the zero frequency limit. 
We find
$r_{\ell\ell}^{(2)}(0)=-1,$
leading to a repulsive contribution from the longitudinal channel when taking
Eqs.~(\ref{F0long}) and
(\ref{rellell}) into account.
Finally, the TM reflection amplitude 
$r_{pp}^{(2)}(0)$ is discussed in
detail in the next section, as its sign
is crucial to determine whether the interaction is attractive or repulsive. 
 
 \vspace{1cm}

\section{TM reflection coefficient in the presence of nonlocal media}

\begin{figure}[h]
    \centering
    \includegraphics[width=.95\linewidth]{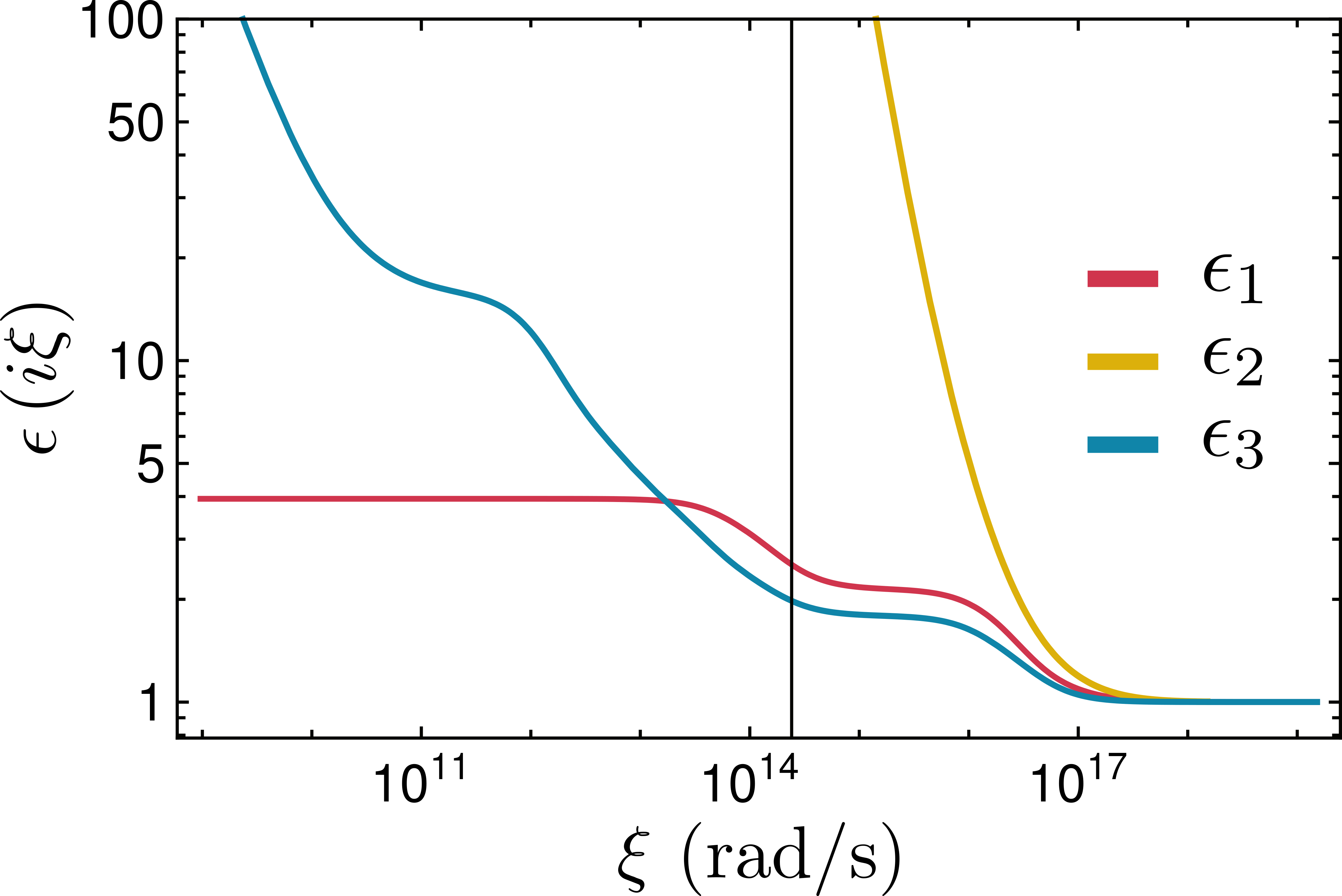}
    \caption{Dielectric functions of SiO$_2$ (red),  and gold (yellow), alongside the transverse dielectric function of salted water (blue), all evaluated along the imaginary frequency axis. The vertical line indicates the first Matsubara frequency $\xi_1$.}
    \label{fig:permitivitties}
\end{figure}

Figure~\ref{fig:permitivitties} shows the behavior of the {transverse} permittivities pertaining to our setup. It is clearly seen that the transverse dielectric function of salted water $\epsilon_3$ lies in between the corresponding values for SiO$_2$ ($\epsilon_1$) and gold 
($\epsilon_2$) for imaginary frequencies below $10^{13}\,{\rm rad/s}.$ Thus, the DLP staircase condition holds in the zero frequency limit, {and not only for the specific materials considered here. Indeed, any configuration with a dieletric surface separated from a metallic one by a layer of salted water also meets the DLP condition at the zero-frequency limit. }

The first nonzero Matsubara frequency is indicated  by a vertical line in Fig.~\ref{fig:permitivitties}, showing that 
 the DLP condition is not met at nonzero Matsubara frequencies. However, one would still find repulsion  
 at sufficiently large distances if nonlocality was neglected. Indeed, 
 {as discussed in the next section,}
 the zero frequency contribution dominates the total Casimir interaction energy at distances $L\gtrsim 100\,{\rm nm}$ for the configuration analyzed here. 
 {This is also the case if the silica half-space is replaced by a  dielectric
 material 
 whose permittivity  also nearly matches the one of water at non-zero Matsubara frequencies
(for instance polystyrene~\cite{Ether2015}, lipids~\cite{Parsegian1971} and other  media of relevance in cell biology~\cite{Spreng2024}). }
 
 In the following, 
 we show that the nonlocal response of salted water breaks the DLP repulsion paradigm leading to attraction at all distances. 
{ More precisely, the nonlocality is encoded in the $K-$dependence of the longitudinal dielectric function $\epsilon_{\ell}(i\xi, K).$ 
Although the reflection coefficients are not directly given in terms of $\epsilon_{\ell}(i\xi, K),$
they depend on $\kappa_{\ell}$ which in turn is determined by the dispersion relation $\epsilon_{\ell}(i\xi, K)=0$ \cite{maia2019scattering}.}

As the reflection coefficient $r_{pp}^{(1)}(0)$ for the SiO$_2$ surface is negative regardless of the model employed for propagation in  salted water, the nature of the interaction is determined by the sign of the reflection coefficient $r_{pp}^{(2)}(0)$ for the  surface of the nanoflake. 
Indeed, the TM contribution
$\mathcal{F}_0^\text{\scriptsize TM}$
is repulsive (attractive) if $r_{pp}^{(2)}(0)$ is  positive (negative)
according to Eq.~(\ref{F0TM}).

If we consider a local model for propagation in salted water, the reflection coefficient is given as the zero-frequency limit of the usual TM Fresnel coefficient~\cite{chew1999waves,born2013principles}: 
\begin{align}\label{local}
         r_{pp}^{(2)}(0)|_\text{\scriptsize{\rm local}}= \lim_{\xi\to 0} \frac{\frac{\omega^2_{p2}}{\gamma_2 \xi}  -\epsilon_{3}(i\xi) }{\frac{\omega^2_{p2}}{\gamma_2 \xi}  + \epsilon_{3}(i\xi) }
\end{align}
The low-frequency limit of $\epsilon_{3}(i\xi)$ is dominated by the 
Drude term in Eq.~(\ref{epsilon1}), then leading to a positive value of $r_{pp}^{(2)}(0)|_\text{\scriptsize{\rm local}}\approx \left(1-2\frac{\omega^2_{p3}}{\omega^2_{p2}}\frac{\gamma_2}{\gamma_3}\right)\sim 1$ since $\omega_{p3}\ll \omega_{p2}.$ The local prescription also provides a positive Fresnel coefficient if one completely disregards the presence of ions in solution and replace $\epsilon_{3}$ by
$\epsilon_b$ in (\ref{local}) (in this case we find 
$ r_{pp}^{(2)}(0)|_\text{\scriptsize{\rm local}}=1$).
As a consequence, the local theory predicts a repulsive zero-frequency contribution with or without ions in solution, in agreement with the DLP repulsion paradigm.

The scattering problem for a plane wave hitting an interface between a nonlocal and a local medium is more involved than the usual case dealt in textbooks \cite{chew1999waves}, as it allows for longitudinal modes and requires an additional boundary condition. 
The TM reflection by the nanoflake surface is analyzed in detail in Appendix A by
following the formalism of Ref.~\cite{maia2019scattering}.
The resulting reflection amplitude in the zero-frequency limit  $r_{pp}^{(2)}(0)$ does not depend on the thickness $d$ of the nanoflake. We use that both $\epsilon_2$
and $\epsilon_3$ behave as $1/\xi$ at low frequencies to find
\begin{align}
     r_{pp}^{(2)}(0)= \lim_{\xi\to 0}\frac{\frac{\omega_{p2}^2}{\gamma_{2}}-\frac{\omega_{p3}^2}{\gamma_{3}\,} -\frac{k}{\kappa_\ell}\frac{1}{\epsilon_{b0}}\frac{\omega_{p2}^2}{\gamma_{2}}\frac{\omega_{p3}^2}{\gamma_{3}}\frac{1}{\xi}}{ \frac{\omega_{p2}^2}{\gamma_{2}}+\frac{\omega_{p3}^2}{\gamma_{3}} +\frac{k}{\kappa_\ell}\frac{1}{\epsilon_{b0}}\frac{\omega_{p2}^2}{\gamma_{2}}\frac{\omega_{p3}^2}{\gamma_{3}}\frac{1}{\xi}}=-1 \, .
     \label{rpplimit}
\end{align}
This result comes from the single requirement that there are ions in solution, which, in turn, allows for a new scattering channel. Indeed, the sign change with respect to the local result (\ref{local}) 
arises from the nonlocal terms in (\ref{rpplimit}) that are proportional to the product $\epsilon_2\epsilon_3\sim 1/\xi^2$ which then dominate in the zero-frequency limit.

When replaced into the scattering formula (\ref{F0TM}), Eq.~(\ref{rpplimit})
leads to an attractive TM contribution that dominates the total zero-frequency term ${\cal F}_0$ as discussed in the next subsection.

\section{Breaking the DLP repulsion paradigm}

As the reflection amplitudes for the two interacting surfaces  $r_{pp}^{(1)}(0)=r_{pp}^{(2)}(0)=-1$ are independent of the dielectric properties of the materials separated by the layer of salted water, we find from (\ref{F0TM}) a universal attractive zero-frequency TM contribution that reads
\begin{eqnarray}\label{F0TMfinal}
 \frac{\mathcal{F}_0^\text{\scriptsize TM}}{A} = -\frac{k_{\rm B} T}{8}  \frac{\zeta(3)}{L^2}\, ,   
\end{eqnarray}
where $\zeta(3)\approx 1.202$ is Ap\'ery's contant. 

The total zero-frequency energy is the sum
(\ref{F0def})
of longitudinal and TM contributions. In order to compare the two contributions, we first note that the product of longitudinal reflections coefficients is bounded as $-1<r_{\ell\ell}^{(1)}r_{\ell\ell}^{(2)}<0$
for our configuration shown in Fig.~\ref{fig:configuration}.
Then, from (\ref{F0long}) and (\ref{F0TMfinal}) we derive the inequality 
\begin{align}
   0<\mathcal{F}_0^\text{\scriptsize long}<\frac{3}{4}\abs{\mathcal{F}_0^\text{\scriptsize TM}}
\end{align}
which implies that the zero-frequency energy is always attractive in spite of the repulsive longitudinal contribution.

We also conclude that the 
total Casimir energy ${\cal F}_{\rm cas}$ is attractive since the dielectric function of SiO$_2$ ($\epsilon_1$) is larger than 
the dielectric function of salted water ($\epsilon_3$) at all nonzero Matsubara frequencies as shown in Fig.~\ref{fig:permitivitties}.

\begin{figure}[h]
    \centering
    \includegraphics[width=0.9\linewidth]{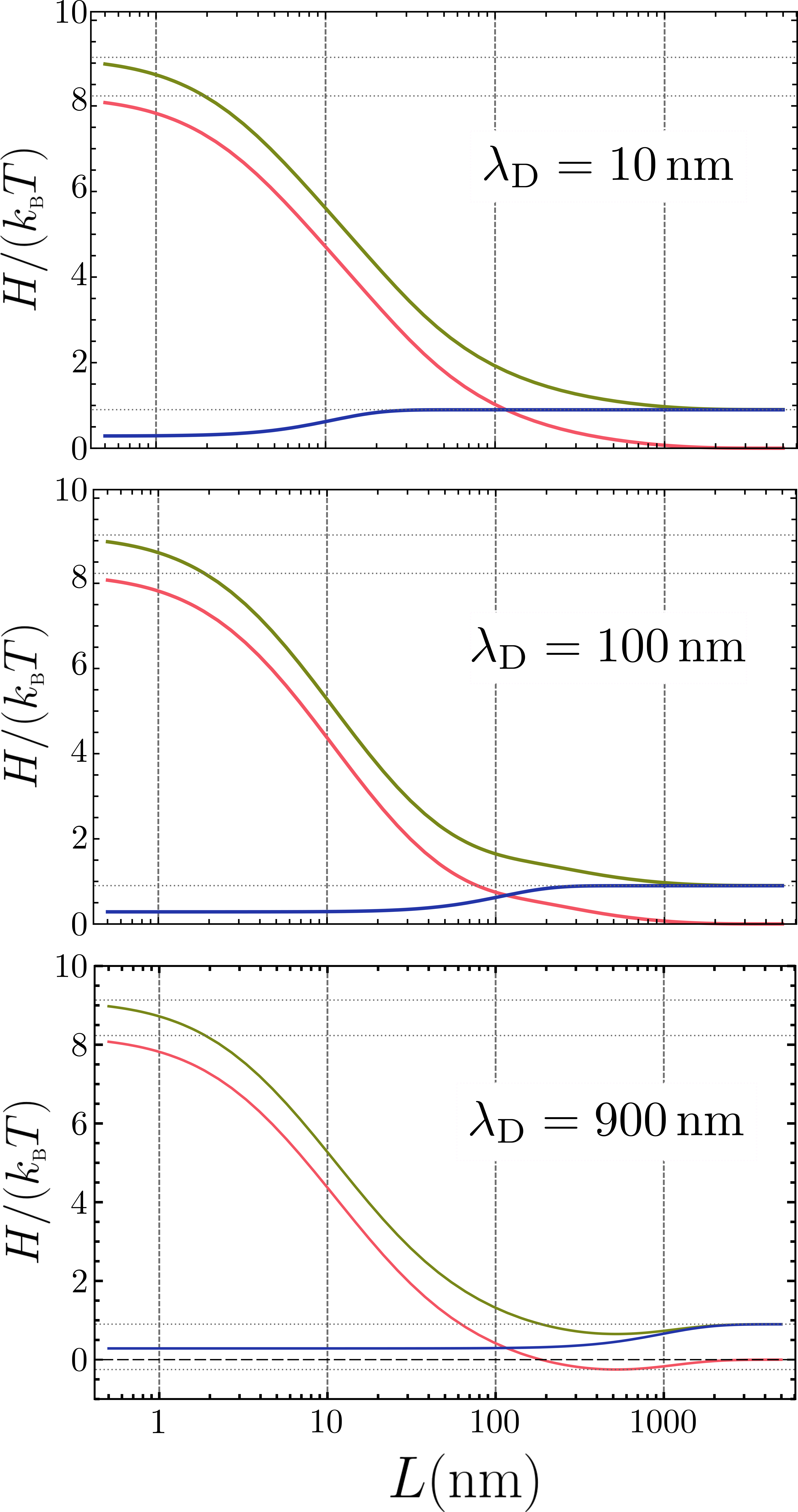}
    \caption{Hamaker function versus distance: total (green), without the TM zero-frequency contribution (red) and zero-frequency contribution including longitudinal and TM channels (blue).   Dotted lines indicate the  asymptotic long-distances limit. From top to bottom, we take the following values for the Debye screening length: $\lambda_{\rm D}=10\,{\rm nm},100\,{\rm nm}$ and $900\,{\rm nm}.$ }
    \label{fig:hamaker}
\end{figure}

We discuss in more detail the different contributions to
${\cal F}_{\rm cas}$ in terms of the
(thermodynamically normalized) Hamaker function
\begin{equation}
    \frac{H}{k_{\rm B} T} = - 12 \pi \frac{L^2}{k_{\rm B} T} \frac{\cal F}{A}
    \label{Hamaker}
\end{equation}
The green line in Fig.~\ref{fig:hamaker} represents  
the total Hamaker function (${\cal F}={\cal F}_{\rm cas}$ in (\ref{Hamaker})) versus distance. The result of subtracting the zero-frequency TM contribution is also shown (red), as well as the zero-frequency contribution (blue), with ${\cal F}={\cal F}_{0}$ in (\ref{Hamaker}).

From top to bottom, the panels in Fig.~\ref{fig:hamaker} correspond to three different Debye lengths: $\lambda_{\rm D} = 10$ nm, $\lambda_{\rm D} = 100$ nm and $\lambda_{\rm D} = 900$ nm. 
{As mentioned in the previous section, the zero-frequency contribution dominates the total free energy for distances $L>100\,{\rm nm}$ in the case of strong or moderate screening.}
 For short distances, we recover the Hamaker constants $H = 8.98 \, {\rm k_{\rm B} T}$ and $H = 8.08 \, {\rm k_{\rm B} T}$ with and without the TM zero-frequency contribution, respectively. For long distances the total Hamaker function goes to $\approx 0.9\, {\rm k_B T}.$
 In the case of  weak screening, 
 the Hamaker function displays a non-monotonic approach to the long-distance asymptotic value, as illustrated by the bottom panel. 
 
It is clear that if the TM zero-frequency contribution was not present, 
the interaction would be repulsive at long distances in the case of weak screening, as illustrated by the red line in the bottom panel of Fig.~\ref{fig:hamaker}. Indeed, 
for sufficiently large values of $\lambda_{\rm D},$ the longitudinal contribution 
${\cal F}_0^{\rm long}$ dominates the contribution from nonzero frequencies at distances smaller than $\lambda_{\rm D}$ and yet larger than the characteristic length scale for electrodynamical retardation. 
However, when the TM zero-frequency contribution is taken into account, the interaction is attractive at all conditions and length scales.

\section{Conclusions}

We investigated the Casimir interaction in the context of the DLP prediction of a repulsive force for a planar three media setup satisfying the so-called staircase condition $\epsilon_1> \epsilon_3>\epsilon_2$ over the relevant frequency range. We showed that the nonlocal response of dissolved ions in the ``medium in the middle" (medium 3) turns the 
otherwise repulsive 
zero-frequency contribution 
into an attractive contribution that 
dominates the total interaction energy at distances larger than $100\,{\rm nm}.$
 This qualitative change in behavior could be important in
the (vertical) trapping of nanoflakes, among other examples. We presented results for a gold nanoflake modeled as a metallic slab of finite thickness, but the breaking of the DLP repulsion paradigm also occurs for a metallic half-space interacting with a dielectric surface across an electrolyte solution. 
{We have considered silica as an example, but the sign change of the Casimir force also holds for other dielectric materials with moderate values of permittivity such as lipids and polystyrene.}

\section*{Acknowledgments}

We thank Fran Gomez and Renan Nunes for helpful discussions. This work was partially supported by Conselho Nacional de Desenvolvimento Cient\'{\i}fico e Tecnol\'ogico (CNPq--Brazil), Coordenaç\~ao de Aperfeiçamento de Pessoal de N\'{\i}vel Superior (CAPES--Brazil),  Instituto Nacional de Ci\^encia e Tecnologia de Fluidos Complexos  (INCT-FCx), and the Research Foundations of the States of Rio de Janeiro (FAPERJ) and S\~ao Paulo (FAPESP).

\appendix

\section{Reflection by a slab immersed in a nonlocal medium}
\label{sec-coefficients}

The appendix aims to prove that 
the zero-frequency contribution in the case of the multilayered system shown in Fig.~\ref{fig:configuration} is the same as in the usual case of 
two semi-infinite half-spaces separated by a layer of salted water.
Indeed, we show that the reflection coefficients $r_{pp}^{(2)}$ and 
$r_{\ell \ell}^{(2)}$
for the interface between salted water and the gold nanoflake are independent of the nanoflake thickness $d$ in the zero-frequency limit. 

\begin{figure}[h]
   \centering
    \includegraphics[width=.95\linewidth]{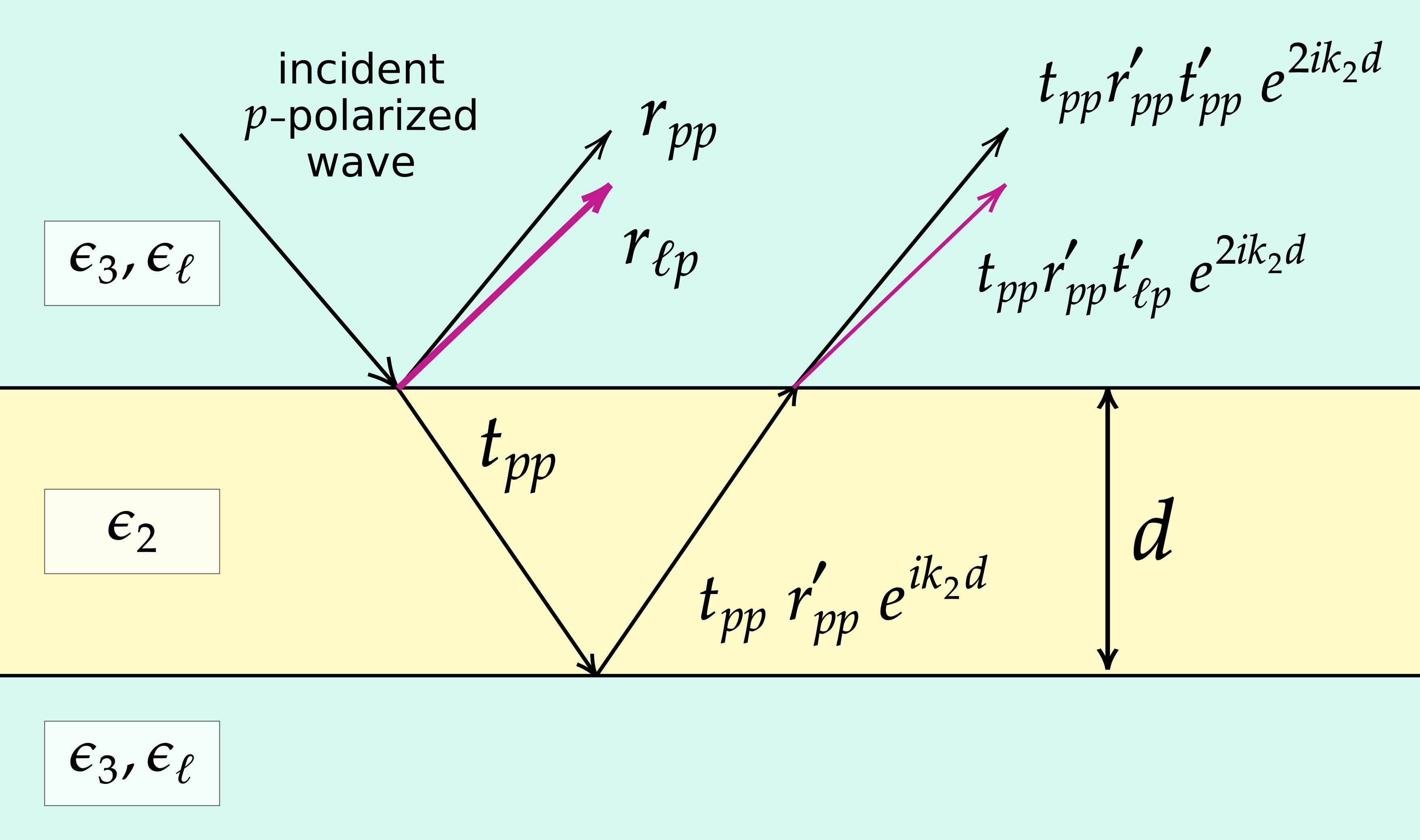}
\caption{Reflection by a slab
of width $d$
and dielectric function $\epsilon_2$
immersed in a nonlocal medium which allows for propagation of longitudinal ($\epsilon_\ell$) and transverse 
($\epsilon_3$) waves. The figure represents the reflection of a $p$-polarized incident wave into transverse and longitudinal  waves given as sums over multiple reflections of $p$-polarized waves within the slab. }
\label{fig-slab-coef}
\end{figure}

As illustrated by Fig. \ref{fig-slab-coef}, the reflection coefficient $r_{\alpha\beta}^{(2)}$ of a 3-layer system (two interfaces) may be obtained by adding the multiple reflections between the two interfaces shown in the figure.
The multiple reflections are written in terms of the single-interface reflection amplitudes $r_{\alpha\beta}$ and
$r_{pp}'$ indicated in Fig.~\ref{fig-slab-coef}.
Here, $\alpha$ and $\beta$ denote the outgoing and incoming polarizations, respectively, which can be either $s$, $p$ or $\ell.$
Multiple reflections give rise to a geometric series, whose summation yields 
\begin{align}
    r_{\alpha\beta}^{(2)}= r_{\alpha\beta}+ \frac{t_{p\beta} r_{pp}^{\prime} t_{\alpha p}^{\prime} \, e^{2ik_{2}d}}{1-{(r_{pp}^{\prime})}^2 e^{2ik_{2}d}}, \label{eq:slab}
\end{align}
 where $k_i$ is the $z-$component of $\textbf{K}_i=(\textbf{k},k_i)$.  Note that the index summation convention is {\it not} in place and 
 that (\ref{eq:slab}) is modified by replacing $p$ by $s$ in its r.-h.-s.
 when considering $\alpha=\beta=s.$

In order to derive the reflection coefficients for a plane interface in the presence of longitudinal modes, we need to solve the Maxwell's equation with boundary conditions for the tangential components of the electric and magnetic fields, and one additional boundary condition \cite{davies1972van},  $\textbf{J}\cdot \hat{z}\mid_{z=0}=0, $ meaning that there is no ``spilling'' of charge across the interfaces. Given the previous discussions \cite{maia2019scattering}, it is fairly easy to see that there is no coupling of TE modes to either TM or longitudinal modes, so that $r_{ss}$
and $r_{ss}^{(2)}$ are the
 standard Fresnel coefficients for a single interface and for a slab, respectively. In contrast, TM and longitudinal modes mix when scattered. 

When considering incoming TM-polarized waves propagating in salted water, the reflection and transmission amplitudes are given by
\cite{maia2019scattering}

\begin{align}
    r_{pp}&=\frac{\epsilon_2 k_3 -\epsilon_3 k_2+\frac{k^2}{k_\ell}\frac{\epsilon_2}{\epsilon_b}(\epsilon_3-\epsilon_b)}{\epsilon_2 k_3 + \epsilon_3 k_2- \frac{k^2}{k_\ell}\frac{\epsilon_2}{\epsilon_b}(\epsilon_3-\epsilon_b)} \\
    r_{\ell p}&= \frac{2\, k_3 \frac{k}{k_\ell}  \frac{\epsilon_2}{\epsilon_b} (\epsilon_3-\epsilon_b)}{\epsilon_2 k_3 + \epsilon_3 k_2- \frac{k^2}{k_\ell}\frac{\epsilon_2}{\epsilon_b}(\epsilon_3-\epsilon_b) } \frac{K_\ell}{K_3} \\
    \label{tpp}
    t_{pp}&=\frac{2 k_3  \epsilon_3 }{\epsilon_2 k_3 + \epsilon_3 k_2 - \frac{k^2}{k_\ell}\frac{\epsilon_2}{\epsilon_b}(\epsilon_3-\epsilon_b)} \frac{K_2}{K_3}\, .
\end{align}
For  incoming longitudinal waves, the amplitudes read~
\begin{align}
    r_{\ell \ell}&=\frac{\epsilon_2 k_3 + \epsilon_3 k_2 + \frac{k^2}{k_\ell}\frac{\epsilon_2}{\epsilon_b}(\epsilon_3-\epsilon_b)}{\epsilon_2 k_3 + \epsilon_3 k_2- \frac{k^2}{k_\ell}\frac{\epsilon_2}{\epsilon_b}(\epsilon_3-\epsilon_b)}\\
    r_{p \ell}&= \frac{2  k \epsilon_2 }{\epsilon_2 k_3 + \epsilon_3 k_2- \frac{k^2}{k_\ell}\frac{\epsilon_2}{\epsilon_b}(\epsilon_3-\epsilon_b) } \frac{K_3}{K_\ell} \\
    \label{tpl}
    t_{p\ell}&= \frac{2  k \epsilon_3 }{\epsilon_2 k_3 + \epsilon_3 k_2- \frac{k^2}{k_\ell}\frac{\epsilon_2}{\epsilon_b}(\epsilon_3-\epsilon_b) } \frac{K_2}{K_\ell}\, .
\end{align}
Finally, Fig.~\ref{fig-slab-coef} shows that we also need 
the amplitudes for an incoming TM-polarized wave propagating in the local medium 2. They are derived by following the method outlined in Ref.~\cite{maia2019scattering}:
\begin{align}
    r_{pp}^\prime&= \frac{{k_3} {\epsilon_2}-{k_2} {\epsilon_3}+\frac{k^2}{k_\ell} \frac{\epsilon_3}{\epsilon_b} ({\epsilon_2}-{\epsilon_b})}{{k_3} {\epsilon_2}+{k_2} {\epsilon_3}- \frac{k^2}{k_\ell} \frac{\epsilon_3}{\epsilon_b}  ({\epsilon_2}-{\epsilon_b})}\\
     \label{tlpprime}
   t_{\ell p}^\prime&= - \frac{2   k_3 \frac{k}{k_\ell} \frac{\epsilon_3}{\epsilon_b} \left(\epsilon_2-\epsilon_b\right)}{{k_3} {\epsilon_2}+{k_2} {\epsilon_3}- \frac{k^2}{k_\ell} \frac{\epsilon_3}{\epsilon_b} 
 ({\epsilon_2}-{\epsilon_b})} \frac{K_\ell}{K_3}\\
 \label{tppprime}
   t_{pp}^\prime&= \frac{2 k_3 \epsilon _3}{{k_3} {\epsilon_2}+{k_2} {\epsilon_3}- \frac{k^2}{k_\ell} \frac{\epsilon_3}{\epsilon_b} 
 ({\epsilon_2}-{\epsilon_b})}\frac{K_2}{K_3}\, .
\end{align}

Combining Eqs.~(\ref{tpp}) and 
(\ref{tpl})
with (\ref{tlpprime}) and (\ref{tppprime}), we obtain the zero-frequency limit
\begin{align}\label{zfl}
    \lim_{\xi\to 0}t_{p\beta} t_{\alpha p}^{\prime} = 0.
\end{align}
When plugging (\ref{zfl}) into
(\ref{eq:slab}), we find that the  reflection coefficient for the slab 
is the same as in the case of a semi-infinite half-space when considering the zero-frequency limit: $r_{\alpha\beta}^{(2)}(0)= r_{\alpha\beta}(0)$.


%

\end{document}